\documentstyle[11pt,paspconf,twoside]{article}

\begin{document}

\title{Summary: Comments on the State of our Subject}
\author{P. J. E. Peebles}
\affil{Joseph Henry Laboratories,
Princeton University, Princeton NJ 08544, USA}

\begin{abstract}
This conference shows the impressive rate of advances in the
observations and theoretical interpretations of large-scale
structure. But to explain my feeling that we may still have a lot
to learn I offer some comments on our sociology, informed by
social construction, in adopting lines of thought that tend not
to follow straightforward readings of what is observed. Thus
galaxies are thought to be affected by environment, not only in
the morphology-density relation, but also as the cause of biased
formation. But the Tully-Fisher and fundamental plane relations
do not allow much room for the contingencies of environment. It is
thought that the galaxy distribution is not likely to be closely 
related to the mass, yet most dynamical analyses indicate that if
the mean density is low then optically selected galaxies are good
mass tracers. The evidence is that at redshift unity the
larger spiral and elliptical galaxies were by and large in 
place, as were a good fraction of the great clusters of galaxies.
These galaxies evolve slowly at $z<1$, yet we hears of rapid
evolution not much before this, in galaxy formation. Perhaps all
these examples are misleading. But, as Kuhn has taught us,
complex interpretations of simple phenomena have been known to be
precursors of paradigm shifts. 
\end{abstract}


\section{Cautionary Remarks}

\begin{quote} {\em The experimental proof of phlogiston seemed
incontrovertible $\ldots$. Indispensable as it was
as a chemical concept, phlogiston as a thing was elusive; it was
widely believed to be the ``least accurately known'' of all
chemical substances or principles, incapable of being isolated
and studied on its own.} In {\it Cavendish the Experimental
Life}, Jungnickel \&\ McCormmach (1999, p. 201) 
\end{quote}

We will remember this as a golden age for our subject, now 
strongly driven by observations as well as theory. But lest we be
remembered for incautiously harboring another phlogiston I
feel some cautionary remarks are in order.

I begin with a lesson to be drawn from the cluster MS1054
(at redshift $z=0.83$). The remarkably detailed observations of 
this cluster are featured in several of the
talks at this conference. The velocity dispersion of the
galaxies, from the spread of redshifts of their spectra, and the
temperature of 
the intracluster plasma, from the X-ray spectrum, yield
consistent measures of a gravitational potential well deep enough
that standard physics says it is capable of lensing
more distant galaxies, and the effect is 
seen (Tran et al. 1999). That is, we have excellent reason to
believe this is a massive cluster of galaxies, characterized in
some detail, and quite similar to nearby ones. The familiar
and sensible reaction is that a viable theory of
structure formation  must predict the existence of a significant
number of rich clusters at redshift $z\sim 1$. But let us 
pause to consider another lesson: this is a vivid illustration of
the success of textbook physics.  

The analysis of the observations of MS1054 extrapolates standard
physics from the length scales on which it was discovered and
tested --- the Solar System and smaller --- to an object some ten
orders of magnitude larger than the Solar System. It is based
on the detection of electromagnetic radiation at X-ray, optical
and (for other objects at similar distances) radio wavelengths
that has been propagating toward us for half the time our 
universe has been expanding, covering a distance comparable to
the relativistic particle horizon (the limiting distance
allowed by causality in the absence of inflation or whatever
finesses the relativistic singular beginning of a 
universe with positive pressure). We are presented with 
a well cross-checked picture of a 
cluster operating according to the physics that was in 
our textbooks well before technology permitted its application on 
such enormous scales of size and distance. One can adduce many
more examples of the success of standard physics, of course.
Physical scientists may seem  to our colleagues in 
sociology a little didactic if not downright arrogant in our
advertisement of textbook physics as a good approximation to
an objective physical reality, but this is a response to
exceedingly strong reinforcement.

We have to be a little more cautious about the lines 
of research we have chosen to follow in seeking to add to
textbook physics; the choice has to be at least in part a social
decision. Hacking (1999) discusses ``sticking
points'' that tend to hang up debates on physical science as a
social construction as opposed to an approach to an external
objective reality.\footnote{All I
know about social construction
is what I read in the  reviews. I have been most influenced by 
Hacking's (1999) ``foreign correspondent" report.} One point is 
contingency: if we can see research paths not taken we
have to credit the path the community chose with an
element of social construction. Another is stability: if the
chosen path leads to unexpected and successful predictions we are 
encouraged to think it has led to an approximation to a physical
reality. On both points nonbaryonic dark matter
is a good example of a social construction.

We did not have to adopt the
hypothesis of nonbaryonic dark matter; our community could 
have have sought to extend Milgrom's modification of Newtonian
dynamics (Brada \&\ Milgrom 1999 and references therein) 
to a modified relativistic cosmology without dark matter. Given
the full attention and energy of our best and brightest, 
and a few free parameters, maybe we could have come up
with a reasonable fit to a reasonable number of the observations. 
It was sensible that most of us chose to stay with textbook
gravity physics unless or until we're forced away from 
it, leaving a few of our company to explore the options. But this
was a community choice of the kind described in social
construction.
 
The dark matter hypothesis has been stable --- apart from the
shift to a nonbaryonic form --- since its 
introduction more than half a century ago (Zwicky 1933), which is
encouraging. Since Zwicky we have learned that (within standard
physics) the more isolated spiral galaxies have 
massive dark halos that are well correlated with the optical
luminosity, that dark matter is the dominant mass in 
groups and clusters of galaxies, and that it's hard to find dark
mass concentrations unaccompanied by optical objects. Nonbaryonic
dark matter plays an an important role in the most widely discussed
and successful model for structure formation, the adiabatic cold
dark matter (aCDM) model, but the tests of this model are not yet
all that tight, and we do not know how well we could have 
done if we had not had this concept. In
short, I don't think we can argue that nonbaryonic cold dark
matter has proved to be considerably more stable than the
phlogistical  chemistry that occupied much of Henry Cavendish's
scientific career (Jungnickel \&\ McCormmach 1999). Lavoisier's
new chemistry --- without caloric --- evolved 
into the standard model because of its experimental successes, 
in one of Kuhn's (1962) paradigm shifts. Nonbaryonic dark matter could
force its way into standard physics on far less extensive grounds
than we have for chemistry, but we do need hard evidence. If the 
precision measurements of the angular distribution of the thermal
cosmic background radiation fit one of the aCDM models now
under discussion, after adjustment of the model parameters to
values that fit the astronomy, it will make believers of most of
us. But we're not there yet.

It's a natural tendency in our community to act as if we knew 
nonbaryonic cold dark matter exists, as real and firmly
established as the cluster MS1054. 
We've seen spectacular runs of successes in physical science,
and it's reasonable enough to hope that we have another one, that
in aCDM we've hit on a good approximation to how structure forms
after only a few (apparently) false starts --- explosions, hot
dark matter, baryonic dark matter, cosmic strings, and textures. 
This is a positive attitude; it leads most of
our community to focus its attention on one line of thought,
pushing it until the idea proves useful or breaks. It is a
little less benign when talking to people in other sciences, or
the media; they can't be expected to factor in our personal
equations. And we must take care not to fool ourselves. In this
connection I would point to a few clouds on the horizon.

\section{Sticking Points}

The theme for this list of four sticking points (following Hacking
1999) in the search for a standard model for structure formation
is an apparent incongruity between simple phenomena and 
complicated interpretations. The complications are inspired by   
the aCDM model. If this model had the
empirical support of quantum mechanics I would have to
agree that these very likely are examples of complex situations
that only happen to look simple.\footnote{If cold fusion were
observed in the laboratory it would mean quantum mechanics has
unambiguously failed. Since this would happen on scales where
quantum mechanics 
has been abundantly tested it is an excellent bet that cold
fusion fails. Here is a case where it makes sense to pay great
attention to what the theory says, while of course bearing in mind the
slight chance that we have something to learn about fundamental
physics on scales where all evidence so far has been that we
already have an excellent approximation to physical reality.} But
we are searching the phenomena for critical tests of the aCDM model. 

The first sticking point is that if the density parameter in
matter that is capable of clustering (thus excluding a term
in the stress-energy tensor that acts like Einstein's
cosmological constant) is low, $\Omega _m=0.25\pm 0.15$, as 
is pretty generally accepted these days, then optically selected
galaxies are good tracers of mass. Most dynamical studies
are consistent with the assumption that the 
dark mass is in galaxy halos with roughly flat rotation curves
and radii of a few hundred kiloparsecs.\footnote{An exception is
in clusters of galaxies, where the dark mass is smoothly
distributed --- maybe a result of tidal stripping --- and $M/L$
is high --- a readily understandable effect of environment.} So
why do we hear so much about biasing, about   
significant differences between the distributions of
galaxies and mass? It is in part a response to
another observation: spirals, ellipticals, and radio galaxies
have different distributions; they cannot all trace the
mass. But another reason is that biasing follows very naturally
from the aCDM model for structure formation.

The preference of early-type galaxies for denser environments is
consistent with a biasing picture, and good analogs happen in
aCDM model numerical simulations. But how do we reconcile the
contingency of environmental influences 
on galaxies with the regularity of the fundamental plane for
ellipticals and the spheroid components of spiral galaxies? A
straightforward reading would be that the spheroids generally
have been only mildly affected by the environmental influences of
tides and mergers. Some spheroids acquired thin disks;
the Tully-Fisher relation says the resulting luminosity
correlates well with the gravity that determines the rotation
curve. A simple inference would be that, after the decision to
acquire a disk, these galaxies also have evolved as island
universes. 

The second point is the issue of what if anything is in the voids
defined by the large galaxies. 
Most known objects avoid them. This is true of dwarf and
irregular galaxies, and of high surface density gas clouds
detected by absorption lines or HI 21~cm emission. Low surface
density galaxies also avoid the dense concentrations of high
surface density ones, maybe because the large tidal fields
disrupt them, but they also respect the voids (Schombert, Pildis,
\&\ Eder 1997). Very low surface density gas clouds are strongly
excluded from the neighborhood of large galaxies, maybe for
similar reasons; they tend to prefer the edges of concentrations
of galaxies (Shull, Penton, \&\ Stocke 1999). The straightforward 
reading is that the voids are pretty close to empty. A
robust prediction of the aCDM model is that the voids contain a
considerable mass fraction. Elegant numerical simulations of
galaxy formation within this model 
indicate galaxies of all types, classified by
their star formation histories within the model, respect common 
voids, in excellent agreement with the observations 
(eg. Cen \&\ Ostriker 1998,  fig. 2). This impressive result must
be taken seriously, but I find a full suspension of disbelief
difficult. In the model simulations the matter in the voids is in 
clumps --- dark matter potential wells --- capable of producing
gas clouds or star clusters. Potential wells with
escape velocities less than about 20~km~s$^{-1}$ cannot
gravitationally hold baryonic matter heated by photoionization by
the intergalactic ionizing radiation, but there is plenty of room
between that and the escape velocity of a large galaxy for the
formation of dwarf or irregular void galaxies. We know galaxies
with low internal velocities can form --- they are present
in great abundance in the neighborhood of large galaxies --- so
why are they so rare in the voids? Could the void matter in the
aCDM model really have almost totally failed to produce
observable objects that would be expected to bear the stigmata of
a troubled youth in a hostile environment? 

The third point is another aspect of biasing, in the relation
between the low order correlation functions 
of galaxies and mass. The two-point function for optically
selected galaxies is quite close to a 
power law, $\xi (r)\propto r^{-\gamma}$, $\gamma = 1.77\pm 0.04$,
over nearly three orders of magnitude of separation, from a
few tens of kiloparsecs to about 10~Mpc 
(when scaled to $H_o=100$ km~s$^{-1}$~Mpc$^{-1}$). The power law
index $\gamma$ changes little back to $z\sim 1$. The 
three- and four-point galaxy functions are less accurately
measured, but are consistent with a simple scale-free clustering
hierarchy on scales less than a few megaparsecs.  
The aCDM models predict that the ratio of galaxy and mass
two-point functions varies with time and separation. 
The models can be adjusted to make the galaxy function a close
approximation to a power law, with index $\gamma$ that is close
to stable back to $z\sim 1$, out of the more complicated
behavior of the mass correlation functions (eg. Pearce et al.
1999). This is impressive but curious: since mass controls
dynamics why would the galaxies considered as mere froth display
the power law regularity?  

At this meeting Adelberger presents striking evidence that the
galaxy and mass two-point functions are quite different at 
redshift $z\sim 3$, as predicted by aCDM model simulations.
We will have to see whether this is a property of galaxies in
general at $z\sim 3$ or an analog of the strong clustering of
present-day radio-selected relative to optically selected
galaxies. If the former, then the galaxies between the
concentrations of Lyman-break systems observed by Adelberger {\it
et al.} have to have formed later than $z=3$, as predicted by
aCDM models. But that leads to the fourth apparent incongruity.

The observations indicate that at redshift $z=1$ the bulk of
the high surface density ellipticals have formed and settled
to their fundamental plane, after correction for evolution
of the star populations, that much the same is true of the
large spiral galaxies, and that there is in addition a population
of more rapidly fading lower luminosity galaxies that maybe end
up as the abundant present-day low luminosity low velocity
dispersion companions of the
giants. At this conference we heard that there is not a lot of
evidence for evolution of the rich clusters of galaxies: at
$z\sim 1$ they tend to be less well relaxed, and maybe they tend
to be less massive, but not by a large factor. One gets the
impression that an astronomer sent back in time to $z=1$, with a
telescope and all technical support, would recognize the 
objects that will evolve, as island universes or close to it,
into what appear in our textbooks of astronomy.
Continuity might suggest the evolution of structure was not
dramatically different in the last factor of two expansion,
from $z=1$ to the present, and in the previous factor of two, 
$z=3$ to $z=1$. 

There is some evidence for this. Prochaska \&\
Wolfe (1998) make a case for the interpretation of the
damped  Lyman~$\alpha$ absorbers (DLAs) as young disk galaxies,
already assembled at $z\sim 3$. Gilmore (1999) and colleagues
find evidence for small scatter in ages of the stars in the old
spheriod population of our galaxy, and indications of a similar
situation in other large galaxies. If the stars formed at 
$z\sim 1$ it would require a most curious synchronization; 
it's easier to think the stars formed early, when the age of
the expanding universe was comparable to the spread in star ages
in the first large generation.
The stars could have formed early but assembled in
galaxies much later, but this late
assembly would require a large collapse factor.\footnote{At the
line of sight velocity  dispersion characteristic of an $L_\ast$
elliptical, $\sigma =150$ km~s$^{-1}$, and at redshift $z=1$, the
mean density within the radius $10h^{-1}$~kpc is about $3\times
10^4$ times the background value.} If there were large collapse
factors in the gravitational assembly of galaxies at $z\sim 1$,
why don't we see large collapse factors at lower
redshift?\footnote{An example would be the collapse of
the Local Group, in a crossing time, all the way to the merging
of the Milky Way and M31. That would require that the relative
motion is close to radial. But numerical solutions indicate the
transverse velocity is comparable to the radial velocity of
approach of these two galaxies (Peebles 1994), a result of the
perturbation by neighboring mass concentrations. If this is so
the Local Group is not going to collapse much further. And
neither are the great clusters, according to the evidence we
heard at this meeting.} There is
clear evidence of merging and accretion 
at $z\la 1$, but as a perturbation to the big picture of the
evolution of galaxies and the central parts of clusters of
galaxies as nearly isolated and stable island universes. If
galaxies were assembled  
early, at $z\gg 1$ it would require more modest and maybe more
reasonable-looking collapse factors. 

As mentioned above, the Lyman break objects do present evidence
for late galaxy formation, and this is in line with the observed
rapid decline in the mean star formation rate from 
$z=1$ to the present (Madau 1999; Steidel et al. 1998). But the
mean star formation rate is close to flat at $1\la z\la 3$, and the
density parameter in hydrogen in DLAs shows a monotonic increase
with increasing redshift, reaching a value comparable to that now
in stars in spirals at $z=3$ (Storrie-Lombardi \&\ Wolfe 1999),
as if the baryons had already been assembled in protogalaxies by
$z=3$. 

The aCDM model prediction of late galaxy formation, peaking
at $z\sim 1$, is not inconsistent with the evidence that
spirals and ellipticals are present at $z\sim 1$, 
and it certainly is not inconsistent with the more confused
observational picture of what happened earlier. But 
it is curious that the proposed rapid evolution of structure at
$1\la z\la 3$ is so different from the mild gravitational
evolution seen in the last factor of two expansion.

These four points show aspects of present-day research in
structure formation that Kuhn (1992) might characterize as
typical of the precursor of a paradigm 
shift. Maybe this is a result of the intense search for empirical
regularities as clues to how  galaxies formed; maybe some of our
sticking points are social constructions rather than true
phenomena. Maybe others are examples of the simple patterns that
can come out of complex physics. And maybe some indicate we still
have things to learn about the physical basis for the evolution
of structure at $z\la 3$. 

\section{The State of Cherished Hypotheses}

Most of us would agree that there are no pressing crises in
cosmology, in the sense that the paradigms have been easily 
adjusted to consistency with the observational advances.
There are clouds on the horizon, that may be only glitches or may
signal changes to come. To organize our thoughts let us consider
how we might react to failures of common hypotheses. 

Nonbaryonic dark matter is hypothetical but well motivated: it
reconciles the dynamical measures of the mean mass density with
the baryon density in the standard model for the origin of
deuterium, and it has simplified the search for a model for
structure formation. Most of us would be quite surprised to lose
nonbaryonic dark matter, though we certainly would be comforted
to see it acquire some color, as an interaction in the laboratory. 

The consensus is that the mass density in matter capable of
clustering is $\Omega _m=0.25\pm 0.15$. The case is not
definitive, and it implies we flourish at a special epoch, but I
think few of us expect to see a return to the Einstein-de~Sitter
model. The case for a mass component that acts like Einstein's
cosmological constant depends on the SNeIa redshift-magnitude
relation and the 
spectrum of angular fluctuations of the thermal cosmic background
radiation. The former is a beautiful but difficult measurement.
The latter is work in progress. A shift to an open model might 
not be traumatic. 

Gravitationally driven hierarchical structure formation is well
observed, in the collapse of the Local 
Group, virgocentric flow, merging of clusters, and merging of
individual galaxies. This might be counted as part of the
established standard model for structure formation (with an
assist from nongravitational processes, as in intracluster
plasma and galactic winds). It is sensible to extrapolate to
the hypothesis of hierarchical assembly of galaxies. Nearby
galaxy merging events offer a reasonable model for what is seen
at high redshift, and this has led to the conclusion that galaxy
merging was a lot more common at $z\sim 1$: perhaps we are seeing
the last stages of hierarchical assembly of the
galaxies. But irregular morphology need not signify youth; the
assembly of the mass concentration in a galaxy, whether by 
hierarchical growth or monolithic collapse, could
happen well before the distribution of light has relaxed to a
textbook galaxy. The many close pairs of galaxies
in the cluster MS1054 could signal merger events about to
happen, or maybe they indicate the cluster has recently
accreted some loose groups that are falling apart. Though I would
be startled to see the failure of the paradigm of gravitational
growth of clustering on relatively large scales, I see a less
strong case on comoving scales $\la 100$ kpc. 

The evidence is that at $z=1$ the large galaxies and the great
clusters of galaxies are not greatly different from today, and
present in comparable numbers. There is a considerably higher
global rate of star formation, maybe because many galaxies have
just formed, or maybe because it took a long time to convert
gas to stars in systems that had been assembled as mass
concentrations much earlier. This meeting featured discussions of
an elegant possible signal of galaxy formation, an increase of
the comoving galaxy clustering length at redshifts approaching
the epoch of first generation of galaxies. There is
observational evidence of this effect, at $z\sim 3$, as predicted
by the aCDM model for structure formation. But the last section
lists reasons why some of us might not be entirely surprised to
see some fundamental adjustments of the picture. 

We have strong faith in the stability of textbook physics, for
good reason. It is not unnatural to have pretty strong faith in
our ideas on how to add to the textbook physics, as the aCDM
model, but sensible to be a little cautious because we are
attempting to draw large conclusions from limited evidence. At
the rate people are improving the observations, and our
understanding of the predictions of the aCDM model, as 
documented in these Proceedings, we may soon learn
whether we have already have a theory for galaxy formation 
that is stable enough to merit promotion to the textbooks, or
whether we are going to have to cast our nets for hypotheses a
little more broadly. Each of us has an opinion on which outcome
would be the more surprising.

\acknowledgments
I am grateful to David Hogg and Bruce Winstein for helpful
discussions. This work was supported in part by Les Rencontres
Internationales de l'IGRAP and the US National Science Foundation.

\end{document}